\documentclass[a4paper,pdftex]{article}
\usepackage{spconf,amsmath,graphicx}

\title{Knowledge Distillation Leveraging Alternative Soft Targets \\from Non-Parallel Qualified Speech Data}
\name{Tohru Nagano, Takashi Fukuda, Gakuto Kurata}
\address{IBM Research AI\\ Chuo-ku Hakozaki, Tokyo, 103-8510, JAPAN}

\newcommand{\wl}{{\it word-level }}
\newcommand{\pl}{{\it phone-level }}
\newcommand{\tl}{{\it state-level }}
\newcommand{\iv}{{\it Interview}}
\newcommand{\cl}{{\it Classroom}}

\begin{document}
\ninept

\maketitle

% ----------------------------------------------------------------------------------------------------
% Abstract
% ----------------------------------------------------------------------------------------------------
\begin{abstract}
This paper describes a novel knowledge distillation framework that leverages acoustically qualified speech data included in an existing training data pool as privileged information. In our proposed framework, a student network is trained with multiple soft targets for each utterance that consist of main soft targets from original speakers' utterance and alternative targets from other speakers' utterances spoken under better acoustic conditions as a secondary view. These qualified utterances from other speakers, used to generate better soft targets, are collected from a qualified data pool by using strict constraints in terms of word/phone/state durations. Our proposed method is a form of target-side data augmentation that creates multiple copies of data with corresponding better soft targets obtained from a qualified data pool. We show in our experiments under acoustic model adaptation settings that the proposed method, exploiting better soft targets obtained from various speakers, can further improve recognition accuracy compared with conventional methods using only soft targets from original speakers.
\end{abstract}


% ----------------------------------------------------------------------------------------------------
% Introduction
% ----------------------------------------------------------------------------------------------------
\vspace{-4ex}

\section{Introduction}

Complex acoustic models often cannot be deployed for real-time decoding of streaming speech data because of the constraints they pose in terms of latency and computation resources. Attempts to tackle this limitation include using compact models trained via knowledge distillation \cite{44873,geras16blendingws,Chan2015,Tang2016,Li2014,Manohar2018,Mosner2019,Ba&Carauana2014}. In the knowledge distillation framework, instead of training models that have reduced computational requirements and improved latency performances directly on hard targets in a single step, training is performed in two separate steps. First, complex teacher acoustic models are trained with a sufficient amount of training data. A compact student acoustic model is then trained on the soft outputs from a teacher network using training criteria that minimize the differences between the student and teacher distributions. This technique has been very successful in various settings -- fully supervised \cite{geras16blendingws}, semi-supervised \cite{Chan2015}, multilingual \cite{Cui2017}, and sequence training \cite{Wong2016} -- for training student networks that perform better than similar models trained from scratch using hard targets.

To improve the accuracy of student networks, more recent work has focused on techniques for leveraging information from multiple teachers with training student networks on an ensemble of teachers \cite{fukuda2017InS}. In addition, as an extension to the standard knowledge distillation framework, generalized knowledge distillation techniques have recently attracted attention as a way of supplementing missing information or compensating for the inferior quality of acoustic features by using privileged information \cite{Vapnik2015}. Several methods related to this generalized knowledge distillation framework with privileged information were previously investigated for Automatic Speech Recognition (ASR) \cite{fukuda2017InS, markov2016, fukuda2019}. In \cite{markov2016}, clean speech is used to generate better soft labels for student networks, while the corresponding noisy speech is used to train noise-robust student acoustic models. In our previous works \cite{fukuda2017InS, fukuda2019}, we investigated the efficacy of using soft labels created by a wideband teacher network as privileged information to improve narrowband student networks.
These previous works demonstrated the potential to improve ASR accuracies on student networks, where low quality features are often fed to the students at the decoding time, by implicitly distilling knowledge from teachers with better quality features as privileged information.

However, such approaches require additional time-aligned parallel speech data to improve their accuracy on student networks.
It is usually difficult to obtain such ideal parallel data for every ASR application because of data collection costs. On the basis of these observations, we hypothesize that leveraging alternative training data from an existing data pool instead of collecting a parallel corpus is key to improving speech recognition accuracy at a low cost. More specifically, we propose a method to create additional or alternative qualified soft labels from audio data which is different from an input to the student model, uttered by different speakers. In our proposed method, we first design a qualified data pool to exploit privileged information from existing speech corpora recorded in acoustically good conditions. Then, privileged soft labels are collected from the qualified data pool by seeking for word candidates paired to the original speaker's utterances, which are spoken in challenging practical scenarios, with word/phone/state-level constraints regarding their durations. By using this scheme, acoustically better quality data can be exploited for better soft label generation without using well-designed parallel corpora. To illustrate the efficacy of our approach, we show how a domain-specific convolutional neural network (CNN) acoustic model can be constructed by a model adaptation technique using privileged soft labels obtained from a qualified data pool, instead of training the network only on single soft labels from original speakers. Experiments have been carried out using a knowledge distillation framework known to be also effective for the domain adaptation of acoustic models in hybrid speech recognition systems \cite{Li2017,8683422,7953145,9003776}. Unlike domain adaptation using single soft labels, which is often done with knowledge distillation, our method can incorporate privileged information corresponding to original output labels from a qualified data pool as an additional view of the data.

\vspace{-1ex}

\begin{figure}[tbp]
  \centering
  \includegraphics[width=\linewidth]{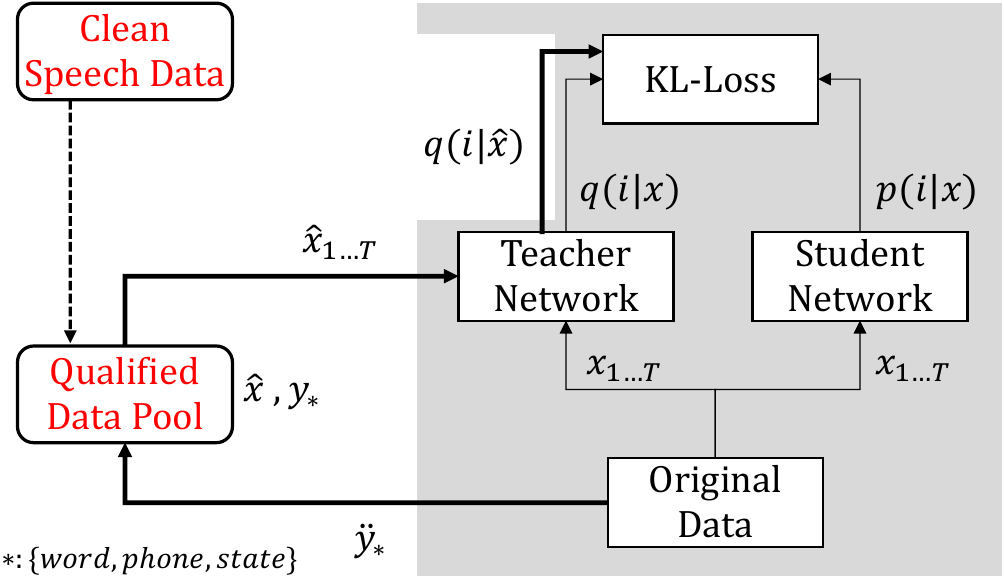}
  \caption{Overview of training process using multiple soft targets. The box with gray background shows the schematic diagram of the conventional knowledge distillation process that uses single soft labels. $\ddot{y}_{*} $ is constraint for extracting alternative speech segments from qualified data pool.
  }
  \label{fig:concept1}
\end{figure}

\begin{figure*}[tbp]
  \centering
  \includegraphics[width=\linewidth]{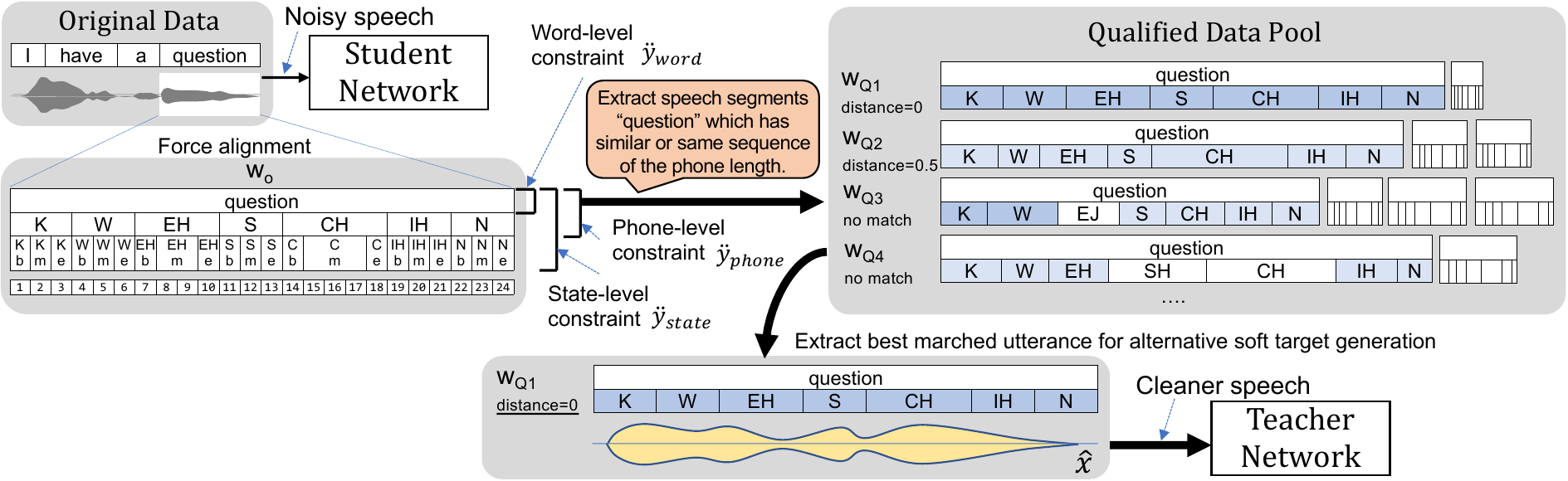}
  \caption{
    For a word ``question'' in the original training data, we select the best matched utterance region ``question'' that has the same word / phone / states sequence.
  }
  \label{fig:concept2}
\end{figure*}

% ----------------------------------------------------------------------------------------------------
% Knowledge Distillation using Non-Parallel Data
% ----------------------------------------------------------------------------------------------------
\section{Knowledge Distillation using Non-Parallel Data \label{sec:distillation}}

\subsection{Training with Multiple Soft Targets}
 In the conventional knowledge distillation framework, training is performed in two separate steps. First, complex teacher neural networks such as bidirectional long short-term memory (LSTM), VGG \cite{Sercu2016}, and residual network (ResNet) \cite{He2016} models are initially trained using hard targets. Compact acoustic models or student networks are then trained on the soft outputs of teachers corresponding to the same speaker's utterances using training criteria that minimizes the differences between the student and teacher distributions (The box with a gray background of the Figure \ref{fig:concept1}). Recently, this knowledge distillation has been extended to use privileged information available only during training \cite{Vapnik2015}. The knowledge distillation training with privileged information is expressed with a soft label $q(i|\hat{\mbox{\boldmath$x$}})$ generated with a better quality feature $\hat{\mbox{\boldmath$x$}}$, which is time-aligned to the degraded quality feature $\mbox{\boldmath$x$}$ as:
\begin{equation}
  {\cal L} (\theta) = -\sum_{i} q(i|\hat{\mbox{\boldmath$x$}})\log p(i|\mbox{\boldmath$x$}),
\end{equation}
where the teacher network is trained with acoustically better quality features $\hat{\mbox{\boldmath$x$}}\in\cal{\hat{X}}$ (or both $\mbox{\boldmath$x$}\in\cal{X}$ and $\hat{\mbox{\boldmath$x$}}\in\cal{\hat{X}}$). $i$ is class label. Instead of using $\mbox{\boldmath$x$}$, the corresponding better feature $\hat{\mbox{\boldmath$x$}}$ is used to generate better soft labels $q(i|\hat{\mbox{\boldmath$x$}})$, while degraded features are used to estimate posteriors $p(i|\mbox{\boldmath$x$})$ for the student network. By training acoustic models with this scheme, the student network can learn privileged knowledge from $\hat{\mbox{\boldmath$x$}}$, which results in creating acoustic models robust to adverse acoustic conditions. The left half of Figure \ref{fig:concept1} shows the process of generating soft labels from the qualified data pool as the privileged knowledge. In the previous literature, for knowledge distillation with privileged information, time-aligned same speakers' utterances are used to estimate $q(i|\hat{\mbox{\boldmath$x$}})$ and $p(i|\mbox{\boldmath$x$})$. In the next section, we describe how privileged soft labels $q(i|\hat{\mbox{\boldmath$x$}})$ are created from a qualified data pool spoken by other speakers under better acoustic conditions.

\subsection{Alternative Target Collection \label{sec:proposed}}
 The key to composing the alternative privileged soft labels is how the better utterances spoken by other speakers that can be used for the distillation training are selected from the qualified data pool. The utterances to be selected include any sentences that are different contents from the original speakers in an utterance level.  Instead, our proposed method for the data selection introduces strict constraints on the durations between original and qualified speech to make knowledge distillation feasible with other speaker's data.

We first design a qualified data pool from existing training data to exploit them as privileged information. From the perspective of privileged information, noiseless speech and speech that has a stable spectral transition such as read speech should be selected as a qualified dataset. In general, speech data recorded in a quiet room by using a close-talk microphone has a higher speech recognition accuracy. Thus, we select cleaner speech data from existing training corpora to make the qualified data pool.

Figure \ref{fig:concept2} illustrates the proposed method for generating alternative privileged soft targets using a qualified data pool. To precisely select alternative speech segments in an utterance, the forced alignment process is first applied to the original data to obtain \tl alignments. During the data collection process in our proposed method, words that have word/phone/state-level alignments in the data pool similar to the word/phone/state labels for original speaker's utterances are selected. For example, in our method, for the word ``question'' in the original speech, the word ``question'' uttered by other speakers in the pool is selected to generate privileged soft labels as a secondary view,unlike the conventional method which obtains the soft label from the same utterance of the same speaker.

The \wl constraint of our method extracts a speech segment where the length of the original speech segment corresponding to the word $l(w_o)$ in the original data is equal to the length of the speech segment $l(w_q)$, where $w_o$ is a word in the original data and $w_q$ is a word in the qualified pool. ``{\footnotesize\texttt{question(24)}}'' in Figure \ref{fig:concept2} is an example of a \wl constraint. The number in parentheses indicates the number of frames. In the \wl constraint, we don't consider the duration in each phoneme between data in the qualified pool and original speaker's data. In contrast, the \pl constraint extracts a speech segment where the sequence of phone and length pairs of word ``question'' in the original data is equal to the sequence of phone and length pairs in the qualified pool. We may consider a margin to tolerate a small difference of durations in each phone. Let each word be a sequence of phones $w = p_{1}, p_{2}, \ldots , p_{N_w} $, where $N_w$ is the number of phones of the word $w$. The average distance of the phones in each word between the original and qualified data is obtained from $L^1$ norm normalized by $N_w$ as $1/N_w^o \sum_{i=1}^{N_w^o}{|l(p_i^o)-l(p_i^q)|}$, where $o$ and $q$ are original data and qualified data respectively. In the \pl constraint with margins, the total number of frames may be different between words in the original data and those found in the qualified pool. In those cases, the lengths of the words in the pool are reorganized by inserting copied soft labels in the context of the preceding and following frames so that the length of the words between the qualified pool and the original data will be the same. ``{\footnotesize\texttt{K(3)-W(3)-EH(4)-S(3)-CH(5)-IH(3)-N(3)}}'' in Figure \ref{fig:concept2} is an example of the \pl constraint. Lastly, the \tl constraint extracts a candidate in the same way as \wl and \pl constraints, but it is performed with more precise analysis based on the \tl alignment information such as ``{\footnotesize\texttt{K(1-1-1)-W(1-1-1)-EH(1-2-1)-S(1-1-1)-CH(1-3-1) -IH(1-1-1)-N(1-1-1)}}'' illustrated in Figure \ref{fig:concept2}.

The best matched words under the constraints mentioned above are used to generate privileged soft labels. When more than two candidates are matched, a word that is closest to the original data with similar meta-attributes is selected. We use these additional alternative soft labels from the qualified data pool, together with the original soft labels obtained from the training data as a multiple view against each word, similar to our previous work \cite{fukuda2017InS}.

% ----------------------------------------------------------------------------------------------------
% Experiments and Results
% ----------------------------------------------------------------------------------------------------
\section{Experiments and Results \label{sec:experiment}}

\subsection{Datasets}
In our experiments on acoustic model adaptation, we used private spontaneous adult and children's speech datasets. The adult spontaneous speech dataset contained a set of Japanese conversations between interviewers and interviewees (\iv\ dataset). The interviews were recorded in a studio with a monaural channel sampled at 16 kHz. We manually labeled the interviewers' and interviewees' speech segments after the recordings. During the interviews, speakers did not follow any instructions about how to speak or what to say, so the recorded data was spontaneous speech with strong accents. Only the subject of each interview, e.g., ``current work'' or ``past work experience,'' was determined prior to the recordings. The \iv\ dataset includes 10.2 hours as adaptation data and 0.5 hours as test data which consists of 55 interviews and 99 speakers.

The children's spontaneous data as the second adaptation data set contained Japanese children's speech privately collected from junior high school students (12 years old) in actual classrooms \footnote[1]{We collected the children's speech data in the lessons that were conducted by Knowledge Constructive Jigsaw (KCJ) method \cite{Shirouzu2016}.}. The recordings were conducted in three classes. The data consisted of a set of discussions between three students. There were 22 students in each class, and 7 groups of 3 students participated in the discussions. The discussion subjects were given by a teacher in advance and included topics such as ``digestion and absorption'' in a science class, ``why we need to cut down trees to protect forestry'' in a social science class, and ``linear functions'' in a math class. Each student wore a headset microphone to record his/her utterances. Each class lasted for 45 minutes, and 10 hours of speech data were generated from the 22 participants in each class. The dataset includes 2.7 hours as adaptation data and 1.9 hours as test data. We manually transcribed the recorded data for our experiments. The average durations of a phone in the \iv\ and \cl\ datasets were 7.58 frames (10 msec per frame) and 7.59 frames, respectively.

\subsection{Qualified Data Pool}
Because these two datasets consist of very spontaneous speech (especially in the children's case) spoken in strong accents, it was more difficult to recognize the utterances than read speech uttered in quiet situations. Our motivation in this paper is to recover such acoustically degraded features of spontaneous speech by exploiting the superior features found in the qualified speech data pool through the knowledge distillation framework. The qualified data pool used to make better soft labels with the proposed method consisted of 350 hours of read speech uttered by several male and female speakers in quiet conditions, which formed a part of the training data for the student CNN and a teacher network (described later). The qualified data pool includes sufficient varieties of context-dependent phones.

\begin{table*}[tbp]
  \caption{
    CERs and matching ratio on two test cases with different selection method.
  }
  \label{tab:results}
  \centering
  \begin{tabular}{p{260pt}|cc|cc}
    \hline
    \multicolumn{1}{c|}{ } & \multicolumn{2}{c|}{\iv\ {\it (Adult)}} & \multicolumn{2}{c}{\cl\ {\it (Child)}}\\
    \multicolumn{1}{c|}{Models} & Matching ratio  & CER & Matching ratio & CER \\
    \hline
    \hline
    \#1 CNN (w/o adaptation)                                                    &  & 21.04 &  & 49.37 \\
    \hline
    \#2 CNN with original soft labels (from original speakers)                  & -- & 19.88 & -- & 45.58 \\
    \hline
    \#3 CNN with original + qualified: equal word length                        & 69.6\% & 19.24 & 70.1\% & 47.18 \\
    \#4 CNN with original + qualified: similar phone length (distance $\le$ 2)  & 77.7\% & 19.35 & 75.8\% & 46.76 \\
    \#5 CNN with original + qualified: equal phone length                       & 24.6\% & {\bf 19.23} & 26.5\% & 45.55 \\
    \#6 CNN with original + qualified: equal HMM state length                   & 9.1\% & 19.51 & 11.5\% & {\bf 44.44} \\
    \hline
  \end{tabular}
\end{table*}

\subsection{Student and Teacher Models}
As a initial model used for a set of adaptation experiments, we trained a CNN-based acoustic model on 2K hours of Japanese broadband speech collected from various data sources. The training corpus included various kinds of in-house datasets: read speech of news articles, conversations between several speakers, far-field speech, command and control utterances, and so on. Children's speech and interview speech data were not included in the training dataset for the initial model.

The CNN-based acoustic model was trained using 40-dimensional log mel-frequency spectra augmented with $\Delta$ and $\Delta \Delta$s as input features. The log mel-frequency spectra were extracted by applying mel-scale integrators to power spectral estimates in short analysis windows (25 ms) of the signal, followed by log transformation. Each frame of speech was also appended with a context of 11 frames after applying a speaker-independent global mean and variance normalization. The CNN system used two convolutional layers with 128 and 256 hidden nodes each in addition to four fully connected layers with 2048 nodes per layer to estimate the posterior probabilities of 9300 output targets. All 128 nodes in the first feature-extracting layer were attached with $9 \times 9$ filters, which were two-dimensionally convoluted with the input log mel-filterbank representation. The second feature-extracting layer, which had 256 nodes, had a similar set of $3 \times 4$ filters that processed the non-linear activations after max pooling from the preceding layer. The non-linear outputs from the second feature-extracting layer were then passed on to the subsequent fully connected layers. All layers used sigmoid nonlinearity. In the assessment of the proposed method, the CNN models were decoded with a tri-gram language model after the initial model was adapted with various settings. The vocabulary contained 300K words and a language model with 300M tri-grams.

The teacher network used in the experiments was a VGG model comprising ten convolutional layers with a max-pooling layer inserted after every three convolutional layers followed by four fully connected layers. All hidden layers had ReLU nonlinearity. Batch normalization was also applied to every fully connected layer. The teacher model was sequence trained after the model was constructed with a cross entropy criterion. The training data for the teacher was the same 2k-hour dataset used for the initial CNN model. Posteriors of the top 50 most likely labels for each prediction of the teacher were then used to adapt the initial CNN model as student networks using the alternative soft label described earlier. The KL-divergence criterion was used for training the student model to minimizing the cross entropy of the soft labels. In our experiments, a part of the 2k-hour training data that was uttered in quiet conditions with a reading style was considered as the qualified data pool, The privileged soft labels created from other speaker's uttrances with our proposed method are used as a secondary view for each input frame of the student network. In addition to soft labels, we also used hard labels by randomly switching the hard and soft labels during the knowledge distillation training.

\subsection{Experimental Results}
The proposed method for collecting alternative soft labels was tested with the various settings for acoustic model adaptation from a well-trained initial CNN model. We evaluated the effectiveness of our proposed method on both the \iv\ and \cl\ test sets containing degraded quality features compared with read speech. As the evaluation metrics, we used the character error rate (CER), as there is ambiguity in Japanese word segmentation.

The experimental results are shown in Table \ref{tab:results}. The initial CNN (\#1) was adapted with the original soft labels to create a baseline adapted model (\#2), which gained 1.16\% and 3.79\% in absolute improvement against the initial CNN model (\#1), respectively. The adapted model (\#2) was created with single soft labels with the original speaker's data obtained from the single teacher model. For the next four adapted student models (\#3 \#4 \#5 \#6), we used alternative soft labels from the qualified data pool obtained by the proposed method, together with the original training data. The difference between the four models was in how the alternative soft labels from the data pool were collected. In our first experiment, student network (\#3) was trained using a mixture of the original soft labels estimated with the same speaker data and alternative soft labels that were matched by the \wl constraint in the qualified data pool. The matching ratio in Table \ref{tab:results} means the proportion of the words found in the qualified data pool with the \wl duration constraint. About 70\% of the words in the original training data were found in the qualified pool when the constraint with word length was used, and these were used in the model adaptation as additional soft labels. Adapted student network \#3 had a 3.2\% relative improvement for the \iv\ dataset over the baseline adapted network (\#2).

Instead of using the \wl constraint as in \#3, adapted models \#4 and \#5 use \pl filtering as more strict constraints. Model \#4 selects the same words from the pool and determines by how much each phone duration of the words differs. It tolerates phone durations of a 2-frame difference when the data is searched for in the pool, while model (\#5) uses only data having exactly the same duration even on the \pl . The model with the weak \pl constraint (\#4) produced a better result for \cl\ compared with the first condition (\#3). The matching ratio for \#4 was larger than that for \#3. The model using the soft labels retrieved under a more severe constraint with the same phone length (\#5) improved both test sets and had a 3.3\% relative improvement for \iv\ over the baseline student systems (\#2), though the matching ratio was much smaller than that of student models \#3 and \#4. The last model (\#6), where the forced alignment of the data was exactly matched by the length of the HMM state (beginning, middle, and end), had a 1.8\% and 2.5\% relative improvement for the \iv\ and \cl\ datasets, respectively, over the baseline network (\#2). These results indicate that the teacher network could successfully integrate privileged knowledge derived from the corresponding soft labels in the qualified data pool into the student network using the original adaptation data.

When comparing the two datasets, we can see that the strongest constraint (\#6) had the biggest accuracy improvement for the children's speech (\cl), and the weaker constraint (\#5) in terms of the matching ratio had the biggest accuracy improvement for the adults' speech (\iv). The accuracy of the baseline for \iv\ was overall better than that for \cl . These results indicate that increasing the number of multiple output views with the weakest constraint leads to better accuracy for speech, which makes it easier for ASR (such as \iv), and selecting carefully qualified speech data leads to better accuracy for acoustically challenging tasks such as children's speech.

% ----------------------------------------------------------------------------------------------------
% Conclusion
% ----------------------------------------------------------------------------------------------------
\section{Conclusion \label{sec:conclusion}}
In this paper, we proposed a novel knowledge distillation framework that leverages better soft labels obtained from an acoustically qualified data as privileged information. We assume that non-parallel speech data with acoustically better conditions can be used as privileged information and can therefore be used in the distillation framework. Our proposed method collects this better quality speech from a qualified data pool by using strict constraints in terms of word/phone/state durations to be used as alternative privileged soft labels. Our practical experiments with different acoustic model adaptation settings showed that the proposed method had a 3.3\% relative improvement for adult spontaneous speech compared with a simple adaptation method based on knowledge distillation and a 2.5\% relative improvement for children's spontaneous speech. These results demonstrate that we can train a better student model using alternative privileged soft labels obtained from other speakers' utterances, compared with a network trained with single soft labels based on the same speaker's speech.

\section{Acknowledgement}
I would like to take this opportunity to thank Prof. Hajime Shirouzu, National Institute for Educational Policy Research, for giving us deep insights and advice about children's collaborative learning.

% ----------------------------------------------------------------------------------------------------
% Bibliography
% ----------------------------------------------------------------------------------------------------
%\newpage
\bibliographystyle{IEEEtran}

\end{document}